\documentclass[aps,prb,twocolumn,showpacs,superscriptaddress]{revtex4}
\usepackage{graphicx,epsf}

\begin{document}
\title{High Dielectric Permittivity in AFe$_{1 / 2}$B$_{1 /
2}$O$_{3}$ Nonferroelectric Perovskite Ceramics (A - Ba, Sr, Ca; B
- Nb, Ta, Sb)}

\author{I. P. Raevski$^{1,2}$ , S. A. Prosandeev$^{1,3}$ , A. S. Bogatin$^{1}$ , M.
A. Malitskaya$^{2}$}

\affiliation{$^{1}$Physics Department, Rostov State University,
344090 Rostov on Don, Russia \\$^{2}$ Research Institute of
Physics, Rostov State University, 344090 Rostov on Don, Russia \\
$^{3}$Ceramics Division, National Institute of Standards and
Technology, Gaithersburg, MD }

\date{\today}

\begin{abstract}

AFe$_{1 / 2}$B$_{1 / 2}$O$_{3 }$(A- Ba, Sr, Ca; B-Nb, Ta, Sb)
ceramics were synthesized and temperature dependencies of the
dielectric permittivity were measured at different frequencies.
The experimental data obtained show very high values of the
dielectric permittivity in a wide temperature interval that is
inherent to so-called high-k materials. The analyses of these data
establish a Maxwell-Wagner mechanism as a main source for the
phenomenon observed.

\end{abstract}

\maketitle

\section{Introduction}

New perovskite-type materials exhibiting very high values of the
dielectric permittivity $\varepsilon '$~ in a very wide
temperature interval discovered during last years open a new
promising possibility for wide applications (the best known
example is CaCu$_{3}$Ti$_{4}$O$_{12}$ which has $\varepsilon '$~
about 10000 which is nearly independent on temperature in the
extremely wide temperature range, 100-500 K, both in ceramic and
single crystal forms \cite{1,2,3,4}). Such characteristics are of
great interest for practice and several possible mechanisms have
been already proposed \cite{2,3,4}. At present it seems that the
origin of the high dielectric permittivity is not intrinsic (as no
anomalies in crystal lattice have been revealed by spectral and
structural studies) but rather extrinsic, most probably- due to
polarization of the Maxwell-Wagner type, but the origin and the
scale of the concomitant electrical and/or chemical
inhomogeneities are still under debate.

Recently high values of $\varepsilon '$~ were reported for ceramic
samples of ternary perovskite BaFe$_{1 / 2}$Nb$_{1 / 2}$O$_{3}$
(BFN) and it was supposed that BFN is a relaxor ferroelectric
\ref{5}. However, a close look into the temperature dependence of
$\varepsilon '$~ obtained in Ref. [\cite{5}] and shown in Fig. 1
(dotted line), as well as the frequency dependencies of both the
real $\varepsilon '$~ and imaginary $\varepsilon "$~ parts of
dielectric permittivity resemble those observed for
CaCu$_{3}$Ti$_{4}$O$_{12 }$. This casts some doubts on the
validity of the consequence made in Ref. [\cite{5}] about the
ferroelectric relaxor nature of the phenomenon observed.

The scope of the present work is the dielectric permittivity
studies of AFe$_{1 / 2}$B$_{1 / 2}$O$_{3}$ (A - Ba, Sr, Ca; B -
Nb, Ta, Sb) in order to elucidate the nature of the high values of
the dielectric permittivity in BFN and similar ceramics. For this
purpose we also studied the electric field and temperature
dependencies of electric current. We found out that all the
ceramics studied exhibit large dielectric permittivity in a wide
temperature interval and that a Maxwell-Wagner mechanism is
responsible for these high values.

\section{Experimental }

The ceramic samples of AFe$_{1 / 2}$B$_{1 / 2}$O$_{3}$~ were
prepared by routine solid-state reaction route. The calculated
quantities of appropriate carbonates and oxides were mixed
thoroughly in an agate mortar in the presence of ethyl alcohol.
Then the synthesis was carried out for 4 hours at 1000-1100 ${
}^\circ\mathrm{C}$. After the synthesis the product of reaction
was crushed and mixed in an agate mortar as described above.
Usually 1wt.{\%} of cubic boron nitride, BN, was added to the
synthesized powder as a sintering aid. The green ceramic samples
were pressed, at 100 MPa, in the form of disks of 5 to 10 mm in
diameter and of 2-4 mm in height. Final sintering was carried out
at $1200-1550^\circ\mathrm{C}$. Pellets were placed on Pt foil.
The density of the
 ceramics obtained was about 90-95 {\%} of theoretical one. The
absence of non-perovskite phases was monitored by X-ray
diffraction. The electrodes for measurements were deposited to the
grinded disk surfaces by rubbing the In-Ga alloy.

The dielectric studies were carried out in the 0.1-100 kHz range
in the course of both heating and cooling at a rate of 2 -3
$^{\circ}\mathrm{C/min}$~ with the aid of the R5083 capacitance
bridge.

In order to avoid the Joule heating of the sample, volt-ampere
characteristics of the semiconducting ceramics were studied by the pulse
method with the aid of oscilloscope using the triangular-shaped voltage
pulses.

\section{Experimental results}

The resistivity $\rho $~ of ceramics obtained depends on the
sintering temperature. The lowest $\rho $~ values $\sim $ 10$^{3
}$- 10$^{4} \mathrm{Ohm}\cdot\mathrm{cm}$~ were observed for
CaFe$_{1 / 2}$Nb$_{1 / 2}$O$_{3}$~ ceramics sintered at
$T>1300^{0}\mathrm{C}$~ while at lower sintering temperatures much
higher $\rho $ values were obtained. Other dielectric
perovskite-type oxides usually have $\rho  \quad \sim $
10$^{7}$-10$^{9} \mathrm{Ohm}^{\cdot}\mathrm{cm}$~ at room
temperature. The type of conductivity was determined by the
Seebeck method for the most conductive samples and was found to be
$n$ for all compounds studied.

The apparent $\rho$~ values of ceramics depend on the measuring
field. Fig. 2 shows the dependences of current density $j$~ on the
electric field $E$~ for three BFN ceramic samples having differing
apparent $\rho $~ values at room temperature. At high $E$'s the
$j$~ on $E$~ dependence is nonlinear and follows the $j \sim
E^{\alpha}$~ relation where exponent $\alpha$~ varies from 3 to 5
for different samples. Such behavior is typical of ceramic
varistors, i.e. ceramics with highly conducting grain bulk and
poorly conducting grain boundary and/or intergranular layers
[6,7]. Small $\alpha$~ values ($\alpha < 1$) at low $E$'s are
usually attributed to carriers hopping along the grain boundaries.
It is worth noting that the same $j$~ vs. $E$~ curves may be
fitted rather well by other types of dependences, e.g. $j\sim \exp
(aE^\nu )$~ with $\nu = 1$~ or $1/2$~ etc. Various mechanisms of
non-linear charge transport across the grain boundary barriers
have been suggested for oxide semiconductors and are still under
debate (see [6,7] and references therein). Discussions of these
mechanisms is out of the scope of the present work. It is only
essential for us that our experiment shows that the ceramics under
study are electrically inhomogeneous.

One more important conclusion follows from the large values of
$E$~ corresponding to the onset of the highly-nonlinear portions
of the $j(E)$~ dependences. This fact implies that the samples
studied contain a large number of high-resistive layers and the
role of possible blocking layer at the sample/electrode interface
is negligible. Qualitatively similar $j(E)$~ dependencies were
observed for all the AFe$_{1 / 2}$B$_{1 / 2}$O$_{3}$~ ceramics
studied provided that the samples have low enough apparent
resistivity.

Fig. 1 displays the $\varepsilon '(T)$~ and $\tan \delta (T)$~
dependences for three BFN samples having differing apparent
resistivity at room temperature (see Table 1). As both
$\varepsilon '$~ and $\tan\delta$~ change by several orders of
magnitude the semilogarithmic scale is used. The $j(E)$~
dependences for BFN-1 and BFN-2 samples are similar to curves 2
and 3 in Fig.2 respectively, while no highly-nonlinear portions
were observed on the $j(E)$~ dependence of the BFN-3 sample in the
available range of $E$~ values (up to 300 V/mm).

All the $\varepsilon '(T)$~ dependences obtained are similar to
those observed for the so-called high-k materials
\cite{1,2,3,4,8}: $\varepsilon '(T)$~ has a step and, above the
temperature of this step, $\varepsilon '$~ are nearly independent
on $T$. Only at high temperatures one can see a small increase of
$\varepsilon '$.

The BFN-1 and BFN-2 samples have very similar $\varepsilon '$~
values and $\varepsilon '(T)$~ dependences; $\tan\delta $~ of
BFN-1 is somewhat higher with respect to BFN-2, especially at room
temperature and above (Fig. 1). These higher $\tan\delta $~ values
may be attributed to higher apparent conductivity of BFN-1.

Both the $\varepsilon '(T)$~ steps and $\tan\delta(T)$~ maxima
shift to higher temperatures as the measuring frequency, $f$,
increases (Figs. 3 and 4). Both the $\varepsilon '(T)$~ and
$\tan\delta(T)$~ maximum temperature, $T_{m}$, dependences on $f$~
plotted in the Arrhenius coordinates are linear (Fig. 4) which is
typical of Debye-type relaxation. The activation energies $\Delta
E_f$~ determined from either of these dependences are very similar
(Fig. 4).

As temperature decreases, $\varepsilon '(T)$~ curves of all the
three BFN samples tend to converge. From our data the
low-temperature limit of $\varepsilon '$~ may be estimated as
40-60. This value seems to be somewhat lower if the measurements
would be extended to lower temperatures and/or higher frequencies.
The high-frequency value of $\varepsilon '$~ for BFN is close to
the estimate $\varepsilon ' \approx 22$~ obtained using the linear
correlation between the $\varepsilon '$~ values measured in the
submillimeter frequency range and tolerance-factor $t =
(R_\mathrm{A} + R_\mathrm{B} ) / \sqrt 2 (R_\mathrm{B} +
R_\mathrm{O} )$, revealed for $\mathrm{A}^{2 + }\mathrm{B}^{3 +
}_{1 / 2}\mathrm{B}^{5 + }_{1 / 2}\mathrm{O}_{3}$~ perovskites
\cite{9} (R$_{\mathrm{A}}$, R$_{\mathrm{B}}$ and R$_{\mathrm{O}}$
stand for the ionic radii of the $\mathrm{A}$, $\mathrm{B}$~ and
$\mathrm{O}$~ ions respectively).

Fig. 5 shows the $\varepsilon '(T)$~ and $\tan\delta (T)$~
dependences for several A$^{2 + }$Fe$_{1 / 2}$B$_{1 / 2}$O$_{3}$~
ceramics. These curves as well as the Arrhenius plots of the
frequency dependence of the $\tan\delta (T)$~ maximum temperature
(Fig.6) are very much alike the corresponding dependences for BFN.
For all the A$^{2 + }$Fe$_{1 / 2}$B$_{1 / 2}$O$_{3}$~ perovskites
studied the low-temperature values of $\varepsilon '$~ are close
to that of BFN ceramics. Note that the temperature at which the
$\varepsilon '(T)$~ step is observed and the activation energy of
relaxation depend on the apparent resistivity values of the
samples. Such behavior is typical of the Maxwell-Wagner relaxation
and below we shall give some additional arguments in favour of
this conclusion.

\section{Discussion}

In the simplest version of the Maxwell-Wagner relaxation model one
can consider alternating slabs having different dielectric
permittivities ($\varepsilon _1 $ and $\varepsilon _2 )$~ and
different conductivities ($\sigma _1 $ and $\sigma _2 )$, so that
the complex permittivities in these slabs are:

\begin{eqnarray}
\label{eq1}
 \varepsilon _1^\ast = \varepsilon _1 - i\sigma _1/\omega  \nonumber \\
 \varepsilon _2^\ast = \varepsilon _2 - i\sigma _2/\omega
\end{eqnarray}
At small conductivities the total dielectric permittivity can be
approximated by

\begin{equation}
\label{eq2}
\varepsilon ^\ast = \frac{L}{l\varepsilon _1^{\ast - 1} + d\varepsilon
_2^{\ast - 1} }
\end{equation}

\noindent where $l$~ and $d$~ are the widths of the first and
second slabs respectively, and $L =l+d$.

Then we consider expression (\ref{eq2}) where, for the sake of
simplicity, only the first slab is conducting. This slab
corresponds to the grain bulk while the nonconducting slab is the
grain boundary. We assume that the temperature dependence of
electroconductivity in the conducting grains is semiconducting

\begin{equation}
\label{eq3} \sigma _1 = \sigma _{10} e^{ - \Delta E_{\sigma} / k_B
T}
\end{equation}

\noindent where $\Delta E_{\sigma}$~ is the electroconductivity
activation energy (this energy can vary for different samples
depending on the degree of the donor-acceptor compensation). This
assumption corresponds to our experimental data on
electroconductivity.

Equating $l$ and $L$~ (this is true for thin interfacial layers
between grains) one can obtain

\begin{equation}
\label{eq4}
\varepsilon ^\ast = \frac{\varepsilon _1 }{a} + \frac{\varepsilon _2
}{a\delta }\frac{1}{1 + i\omega \tau }
\end{equation}

\noindent where $a = 1 + \delta \varepsilon _1 / \varepsilon _2 $,
$\tau = \varepsilon _2 a / \delta \sigma _1 $, and $\delta$~ being
the relative width of the interfacial layers $\delta = d / L$. The
frequency dependence of the dielectric permittivity can be
observed if only $a\delta < < \varepsilon _2$. This can happen in
the case if the nonconductive layers are very thin that
corresponds to our assumption about the grain boundaries, and if
the dielectric permittivity of the conductive region is not very
large with respect to the nonconductive one, and this also
corresponds to our data showing comparatively (with
ferroelectrics) low values of the permittivity at low temperatures
and large frequences.

At zero frequency one can find

\begin{equation}
\label{eq5}
\varepsilon ' = L\varepsilon _2 / d
\end{equation}

\noindent that is the dielectric permittivity is controlled by the
thin nonconducting layer. In the case of ceramics considered in
the present paper this can be boundaries of the grains. For
example, if the grain is about 10 $\mu $m, the boundary region is
about 100 A, and $\varepsilon _2 = 20$~ then $\varepsilon ' =
20000$. This can explain the high values of the dielectric
permittivity in BFN. For instance, at $\varepsilon _1 = 40$, one
can obtain the high frequency limit $\varepsilon ' = 40$~ which
corresponds to our experimental data.

It follows from (\ref{eq4}) that the temperature dependence of the relaxation time
$\tau $ can be described by the Arrhenius law, in which the barrier height
coincides with the electroconductivity activation energy in the grain's
bulk:

\begin{equation}
\label{eq6} \tau = \tau _0 e^{\Delta E_{\sigma} / k_B T}
\end{equation}

\noindent where $\tau _0 = a / \delta \sigma _{10} $. This fact
corresponds to the experimental data showing a close relation of
the electroconductivity in the bulk of the grains to the
relaxation time (see Section 3). Due to temperature dependence
(\ref{eq6}) the real part of the dielectric permittivity should
increase at a temperature determined by the relation $\omega \tau
\approx 1$~ and at larger temperatures it should saturate that is
in agreement with experiment. The greater is the frequency, the
larger is the temperature where the real part of the permittivity
becomes rapidly growing and the imaginary part has a maximum.
These data are in excellent agreement with the experiment
performed.

The microscopic model we use bases on the assumption that the
high-temperature treatment of the ceramics under study results in
the creation of oxygen vacancies which have weakly bound
electrons. The semiconductor conductivity connected with these
electrons contributes into electric polarization and results in
the appearance of a step in $\varepsilon '$~ and a maximum in
$\varepsilon ''$.

We assumed that the interfacial layers at the grain boundaries are
nonconducting (or less conducting) while the bulk of the grains is
conducting. Just this fact together with the relatively small total width of
the grain boundaries in the direction of the field provides the large values
of the dielectric permittivity in the ceramics studied.

Both the resistivity dependence on sintering temperature and
varistor-like $j(E)$ characteristics of the AFe$_{1 / 2}$B$_{1 /
2}$O$_{3}$~ ceramics seem to be due to partial reduction of these
compounds at high temperatures (where the partial O$_{2}$~
pressure in air is lower than the equilibrium partial O$_{2}$~
pressure in oxide). This assumption is consistent with the
$n$-type of conductivity in all the ceramics studied. In the
course of cooling reoxidation takes place. As the oxygen diffusion
coefficient at grain boundaries is much larger than in the bulk,
the degree of oxidation and subsequent resistivity is higher in
the grain boundary layers than in the grain bulk.

As one can see from Table 1, the activation energies of the
relaxation frequency are several times lower than the activation
energies extracted from the temperature dependence of $\rho$. The
former energy coincides with the activation energy of grain bulk
conductivity (see expression (\ref{eq6})) while the latter is
determined by the grain boundary barrier height. The same
situation is observed in CaCu$_{3}$Ti$_{4}$O$_{12}$~ ceramics
where the resistivities of grain bulk and grain boundaries have
been directly determined via the impedance spectroscopy [4]. It is
worth noting that the activation energies observed for the
relaxation time have the same order of magnitude as the activation
energies of the first (hundredths of eV) and the second (tenths of
eV) levels of the oxygen vacancy in perovskites [10,11].

Similar effects, i.e. a crucial dependence of resistivity on
sintering temperature and the formation of grain boundary barrier
layers due to reoxidation have been observed in related compounds
PbFe$_{1 / 2}$Nb$_{1 / 2}$O$_{3 }$~ and PbFe$_{1 / 2}$Ta$_{1 /
2}$O$_{3 }$~\cite{12,13,14}. Both these perovskites are
ferroelectrics and the presence of grain boundary barriers in a
ceramic sample can be monitored by studying the positive
temperature coefficient of resistivity (PTCR) effect, an anomalous
decrease of resistivity on cooling below the ferroelectric Curie
point, $T_{C}$. In the paraelectric phase the properties of these
materials at low frequencies are determined mainly by grain
boundaries: the extremely high permittivity values (though
depending on temperature due to the Curie-Weiss behavior of
$\varepsilon '$~ typical of ferroelectrics), varistor effect, and
high apparent resistivity exhibiting large activation energy were
observed. Below $T_{C}$, the grain boundary barriers are screened,
more or less effectively, due to onset of the spontaneous
polarization. This results in low resistivity values characterized
by small activation energies which are determined by the grain
bulk and relatively small residual grain boundary barriers.

The average grain size in all the BFN ceramics studied was
approximately the same (a few $\mu $m) and the increase of the
$E$~ values corresponding to the onset of the highly-nonlinear
portions on the $j(E)$~ curves of the samples with lower apparent
conductivity is likely to be caused by the increase of the
thickness of the poorly conducting grain boundary layers. This
assumption correlates well (taking into account expression
(\ref{eq5})) with experimentally observed lower values of
$\varepsilon '$~ for the BFN-3 sample as compared with more
conductive BFN-2 and BFN-1.

It follows from expressions (\ref{eq4}) and (\ref{eq6}) that the
temperature behavior of $\tan \delta $~ can have whether a maximum
(at low frequency or at high conductivity) or a step (at large
frequency or at low conductivity). There can be also intermediate
behavior with a step and a maximum just at the step. This
corresponds to the experiment performed as it is seen from Figs. 1
and 6.

Fig. 7 presents the fit of expression (\ref{eq4}) to experimental data obtained for
BFN2. The activation energy of the relaxation process obtained from this fit
is 0.09 eV that is in excellent agreement with the experimental data
obtained from the Arrhenius plot.

We want to stress here that the Debye-type relaxation described
above differs from the usual Debye relaxation connected with the
reversal of local dipoles appreciably. In the case of the local
dipoles $\varepsilon \left( {\omega = 0} \right)\sim 1 / T$~ and
this explains the vanishing of the relaxation contribution to the
dielectric permittivity at high temperatures. The real part of the
dielectric permittivity in this case has a maximum and the
temperature width of the relaxation anomaly is usually not very
large. We expect such a situation for low-conducting samples where
the conducting regions are so small that one cannot substitute the
complex system by a series of capacitors. Rather simple arithmetic
averaging of dielectric permittivity should be done. In this case
one immediately obtains a Lanjevin function describing
polarization, and, at zero measuring field, the permittivity is
$\chi _0 \approx nd^2e / 6k_B T$~ where $d$~ is the width of the
polar region along the field, $e$~ is the electron charge, and
$n$~ is the oxygen vacancy concentration. Notice that this result
is obtained not because of the averaging of polarization over the
possible dipole directions in 3D space as it is true for local
dipoles but rather because of averaging of polarization with
respect to the displacement of an electron from the oxygen vacancy
along the field. The real part of the dielectric permittivity has
a maximum in this case like in the case of local dipoles and the
susceptibility decreases with temperature above the maximum as
1/$T$.

We have considered a monodispersive Debye model. However it can be easily
generalized by considering a distribution function of the relaxation times.
There can be a few relaxation processes connected with different electron's
states. They can lead to different maximums or steps in the imaginary part
of permittivity in different temperature intervals. The spatial dependence
of the parameters of the model can be taken into account by introducing a
distribution function for the relaxation times and energy barriers.

We want to discuss here a possibility of the existence of a
Maxwell-Wagner relaxation in single crystals. Indeed, such
relaxation was observed in single crystals of
CaCu$_{3}$Ti$_{4}$O$_{12 }$[1,2] and ErFe$_{2}$O$_{4}$~ \cite{8}.
In principle this effect could be due to electrodes and a
depletion region near them. In our case we used InGa electrodes
which do not produce such an effect in the case of the n-type
conductivity found in our samples. Another possibility can be
connected with the fact that all high-k materials contain
transition metal ions with open shells like Fe (in our case and in
ErFe$_{2}$O$_{4})$ or Cu (in CaCu$_{3}$Ti$_{4}$O$_{12})$. Below we
will discus a few of possible consequences of this fact.

The oxygen vacancies are comparatively easily created at the
transition metals in perovskites and they are ordered \cite{15}.
This can result in the appearance of reduced regions in the bulk
at the transition metals and these regions can organize clusters
and participate in the conductivity. The ordered arrays of the
vacancies can play the role of conducting layers.

The electronic band structure of AFe$_{1 / 2}$B$_{1 / 2}$O$_{3 }$~
(A- Ba,Sr,Ca; B-Nb,Ta,Sb) is very different in comparison with the
electronic structure of ABO$_{3}$~ perovskites \cite{16}. Indeed
the width of conduction bands in the ABO$_{3}$~ crystals is mainly
determined by the covalent bonding of the $nd$~ metal and O2$p$~
oxygen electronic states. At the Fe ions the covalent bonding is
reduced due to the localized character of the Fe $3d$-states. This
effect should decrease the width of the conduction bands [16] and
produce some percolative clusters made of conducting states. The
electrons from the vacancies caught by these clusters can
participate in charge transfer within these clusters. The charges
of Fe$^{3 + }$~ and B$^{5 + }$~ (B - Nb, Ta, Sb) are very
different, and this also can lead to the appearance of clusters of
electronic conduction states having decreased energy. One can
consider conductivity percolation in samples after different
thermal treatment: for well conductive samples the percolative
clusters are big while for less conductive clusters they are
smaller.

Finally, one can consider electron's jumps over the M-O bands (M =
Cu or Fe). Due to a small oxygen vacancy concentration this
conductivity is prevented by the strong electron correlation on
the metal ions. As a result the electrons can be caught by some
finite regions having random barriers which are higher than the
barriers inside these volumes.

In Ref. [\cite{8}] the authors proposed the appearance of
nonconducting (or less conducting) layers due to a magnetic phase
transition and following domain formations: the domain walls can
play the role of the nonconducting dielectric layers. One can
extend this idea by considering also antiferroelectric or elastic
phase transitions.

Studies of the electron doping on the dielectric properties of
perovskite materials were carried out in pioneering papers
\cite{19,20,21}. It was not ambiguously shown that electrons and
polarons can contribute much to the dielectric permittivity of
perovskites and, at some conditions, produse a temprature maximums
of the dielectric permittivity. From our study it follows that
such "polaronic-type" relaxation is possible in the case when the
polar regions where the electron concentration is enhanced for
some reasons are not very large. In this case these regions show
the same characteristics as local orientable dipoles in relaxors.

\section{Summary}

Experiments performed have shown that ceramics AFe$_{1 / 2}$B$_{1
/ 2}$O$_{3 }$~ (A- Ba, Sr, Ca; B - Nb, Ta, Sb) have similar main
features in the temperature dependence of dielectric permittivity:
$\varepsilon '$~ has a large step in the interval between
$-100^\circ\mathrm{C}$~ and $100^\circ\mathrm{C}$, the magnitude
of $\varepsilon '$~ is low below the step (less then 10$^{2})$~
and large above the step (of the order of 10$^{3 }$- 10$^{4})$;
the large value of $\varepsilon '$~ remains above the step in the
whole temperature interval studied; $\tan\delta $~ has a maximum
or also a step with a small maximum; $\varepsilon '$~ and
$\tan\delta $~ depend on frequency: the respective step and
maximum are shifted to higher temperatures at larger frequencies,
while the magnitude of the permittivity at the maximal value of
the step does not change much. All these features were found to be
influenced by the conductivity of the samples: in the samples with
higher conductivity the step lays at lower temperatures and the
maximal value of $\varepsilon '$~ is higher.

We propose that the phenomenon of the high values of the
dielectric permittivity in ceramics AFe$_{1 / 2}$B$_{1 / 2}$O$_{3
}$~ (A - Ba, Sr, Ca; B - Nb, Ta, Sb) in a wide temperature
interval is due to the Maxwell-Wagner relaxation. We fitted an
expression of this model to the experimental data and found
reasonable agreement between the theory and experiment. It seems
that a $\varepsilon '(T)$~ step-like anomaly reported for SrFe$_{1
/ 2}$Sb$_{1 / 2}$O$_{3 }$~ at $ \approx  220 \mathrm{K}$~ and
attributed to the presence of an antiferroelectric phase
transition \cite{17} is due to the Maxwell-Wagner relaxation. The
same seems to be the case for SrTiO$_{3}$:Bi \cite{18} as well as
for CaCu$_{3}$Ti$_{4}$O$_{12}$ \cite{1,2,3,4}.

\section{Acknowledgements}

This work was partially supported by Russian Foundation for Basic
Research (Grants 01-03-33119 and 01-02-16029).

\newpage
\section*{Figure captions}

Fig. 1. Temperature dependencies of $\varepsilon '$~ (upper
curves) and tan$\delta $~ (lower curves) measured at 1 kHz for
three BFN1 (1), BFN2 (2), and BFN3 (3) (see notations in Table 1).
The dashed line shows the $\varepsilon '(T)$~ dependence for a BFN
ceramics deduced from Ref. [\cite{5}].

Fig. 2. Typical logarithmic plots of $j(E)$~ for BFN ceramic
samples having different apparent resistivity at room temperature.
The upper curve corresponds to a more conductive sumple with $\rho
\sim 10^7 \mathrm{Om \cdot cm}$~ while the lower curve is for a
less conductive sample with $\rho \sim 10^9 \mathrm{Om \cdot cm}$.
The dashed line shows an Ohm-type behavior.

Fig. 3. Temperature dependencies of $\varepsilon '$~ (a),
tan$\delta $~ (b) and $\varepsilon ''$~ (c) for BFN2 (filled
symbols) and BFN3 (open symbols) measured at different
frequencies.

Fig. 4. The Arrhenius plots of the frequency dependence of the
tan$\delta  (T)$~ maximum temperature (filled symbols) and
$\varepsilon ''(T)$~ maximum temperature (open symbols) for BFN-2
and BFN-3 ceramic samples having differing apparent resistivity
values at room temperature. Symbols are the experimental points
and solid lines are the least-squares straight-line fits.

Fig. 5. Temperature dependencies of $\varepsilon '$~ (a) and
$\tan\delta $~ (b) measured at 1 kHz for BaFe$_{1 / 2}$Ta$_{1 /
2}$O$_{3}$~ (1), SrFe$_{1 / 2}$Nb$_{1 / 2}$O$_{3}$~ (2), CaFe$_{1
/ 2}$Nb$_{1 / 2}$O$_{3}$~ (3), and SrFe$_{1 / 2}$Sb$_{1 /
2}$O$_{3}$~ (4) ceramics.

Fig. 6. The Arrhenius plots of the frequency dependence of the
tan$\delta  (T)$~ maximum temperature for BaFe$_{1 / 2}$Ta$_{1 /
2}$O$_{3}$~ (1), SrFe$_{1 / 2}$Nb$_{1 / 2}$O$_{3}$~ (2), CaFe$_{1
/ 2}$Nb$_{1 / 2}$O$_{3}$~ (3), and SrFe$_{1 / 2}$Sb$_{1 /
2}$O$_{3}$~ (4) ceramics. Symbols are the experimental points and
solid lines are the least-squares straight-line fits

Fig. 7. The fit of the Maxwell-Wagner theory to experiment on
BFN-2.

\begin{table}[t]
\caption{Apparent resistivity, $\rho$, at room temperature, and
activation energies $\Delta $E$_{f}$~ and $\Delta $E$_{\rho }$~
for AFe$_{1 / 2}$B$_{1 / 2}$O$_{3}$ ceramics and
CaCu$_{3}$Ti$_{4}$O$_{12}$~ \cite{4}}
\begin{tabular}{c|ccc}
 \hline Compound& $\rho [\mathrm{Ohm\cdot cm}]$& $\Delta E_{f}$ [eV]& $\Delta
E_{\rho } [\mathrm{eV}]$ \\ \hline BaFe$_{1 / 2}$Nb$_{1 /
2}$O$_{3}$ (BFN-1)& $\sim 10^{7}$& 0.08& 0.3
\\ \hline BaFe$_{1 / 2}$Nb$_{1 / 2}$O$_{3}$ (BFN-2)& $\sim
10^{8}$& 0,08& 0.3 \\ \hline BaFe$_{1 / 2}$Nb$_{1 / 2}$O$_{3}$
(BFN-3)& $\sim 10^{9}$& 0.35& 0.55 \\ \hline BaFe$_{1 / 2}$Nb$_{1
/ 2}$O$_{3}$ [ 5 ]& $\sim 10^{8}$& 0.12&
 \\
\hline BaFe$_{1 / 2}$Ta$_{1 / 2}$O$_{3}$ & $\sim 10^{8}$& 0.08&
0.4 \\ \hline SrFe$_{1 / 2}$Nb$_{1 / 2}$O$_{3}$ & $\sim 10^{8}$&
0.16& 0.4 \\ \hline CaFe$_{1 / 2}$Nb$_{1 / 2}$O$_{3}$ & $\sim
10^{9}$& 0.3& 0.7 \\ \hline SrFe$_{1 / 2}$Sb$_{1 / 2}$O$_{3}$ &
$\sim 10^{9}$& 0.4& 0.7 \\ \hline CaCu$_{3}$Ti$_{4}$O$_{12 }$~
\cite{4}& & 0.08& 0.6 \\ \hline
\end{tabular}
\label{tab1}
\end{table}

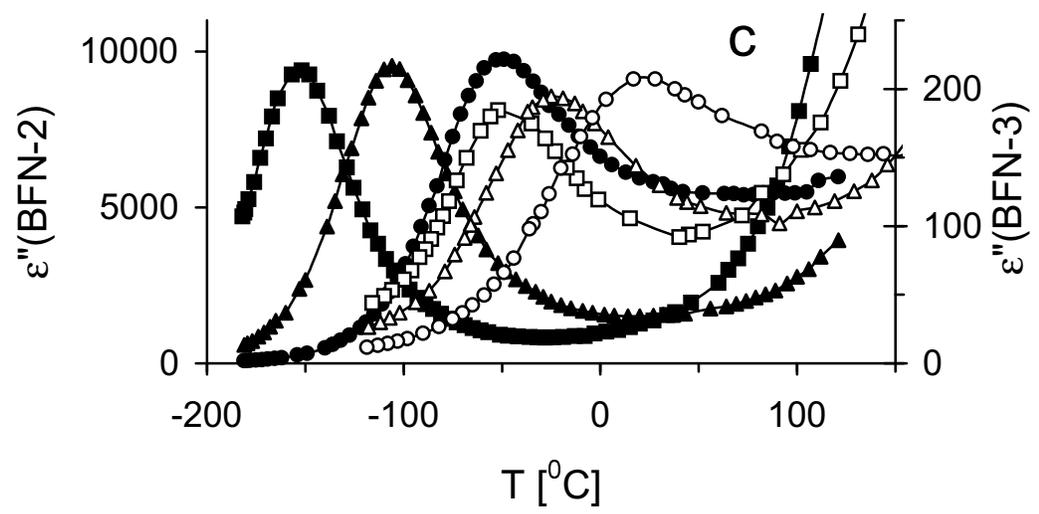
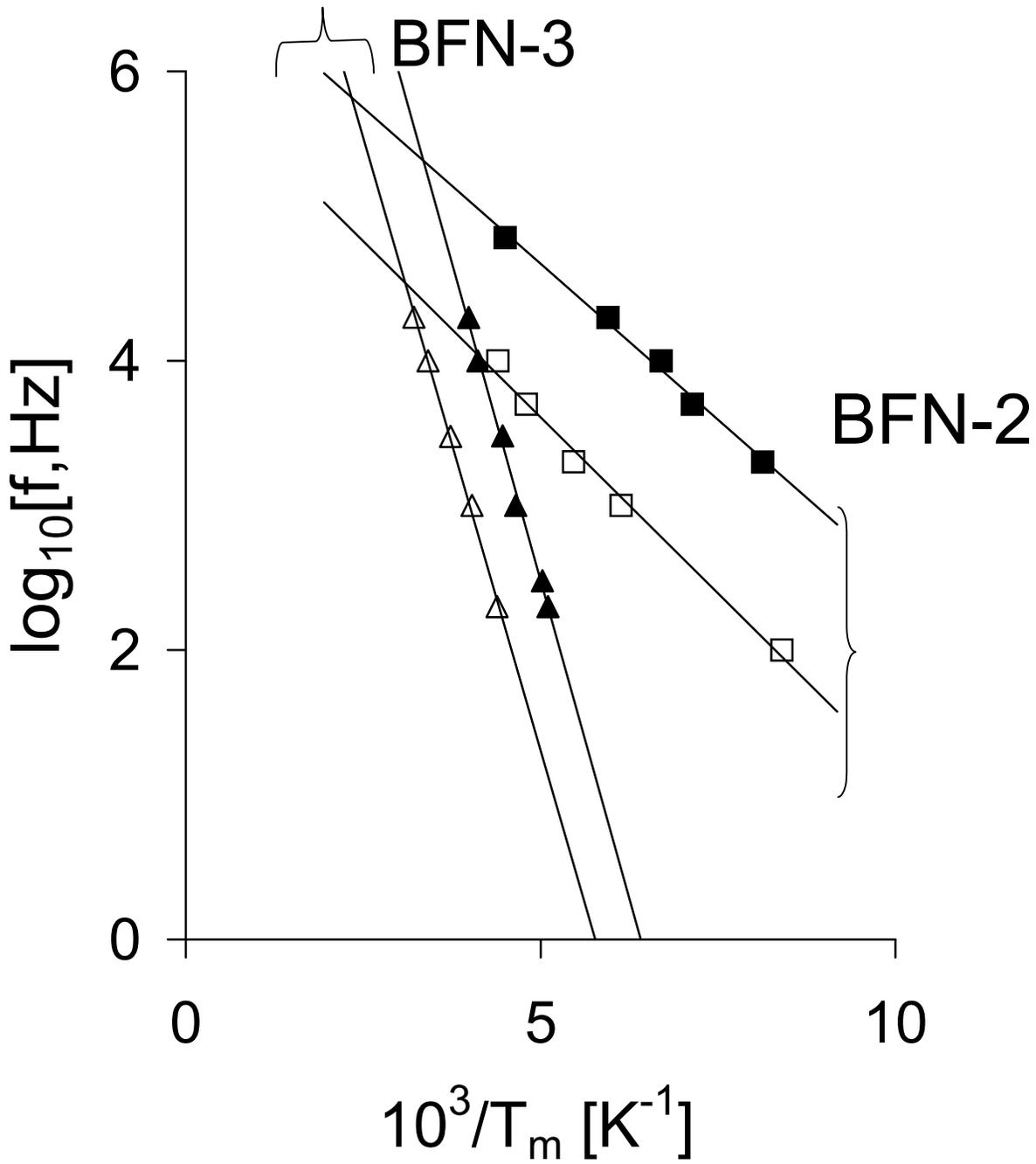
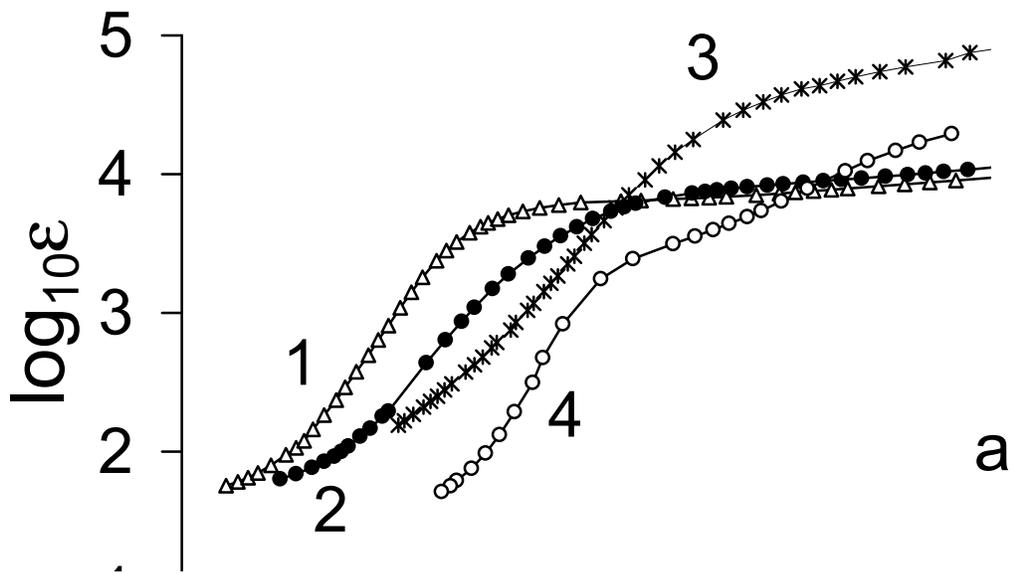
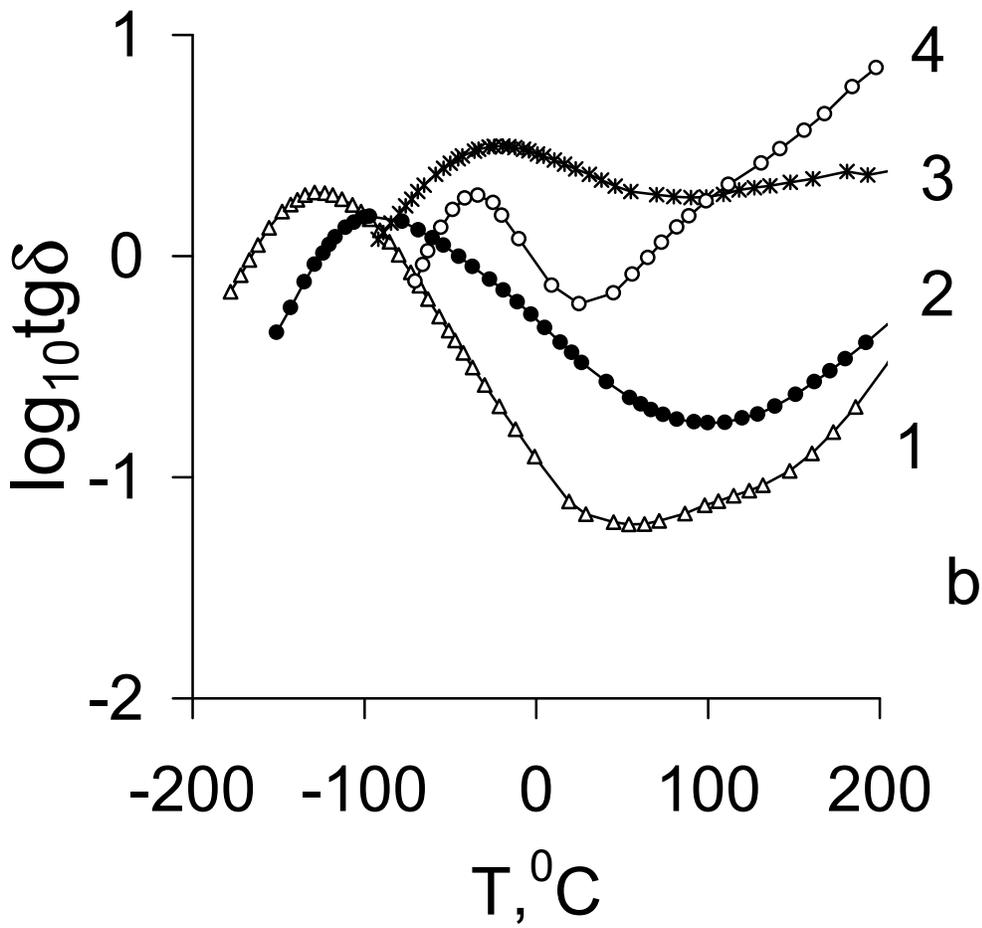
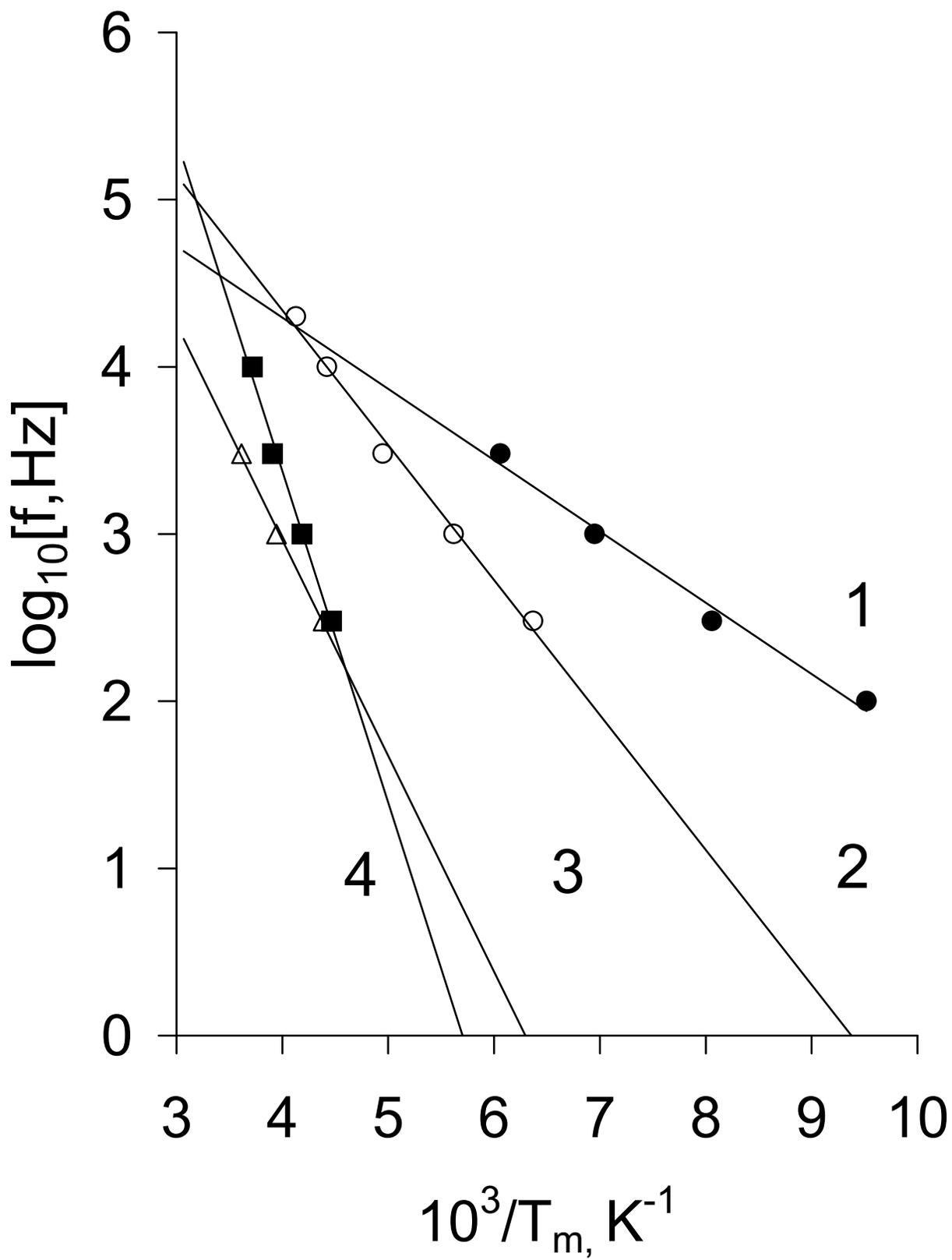
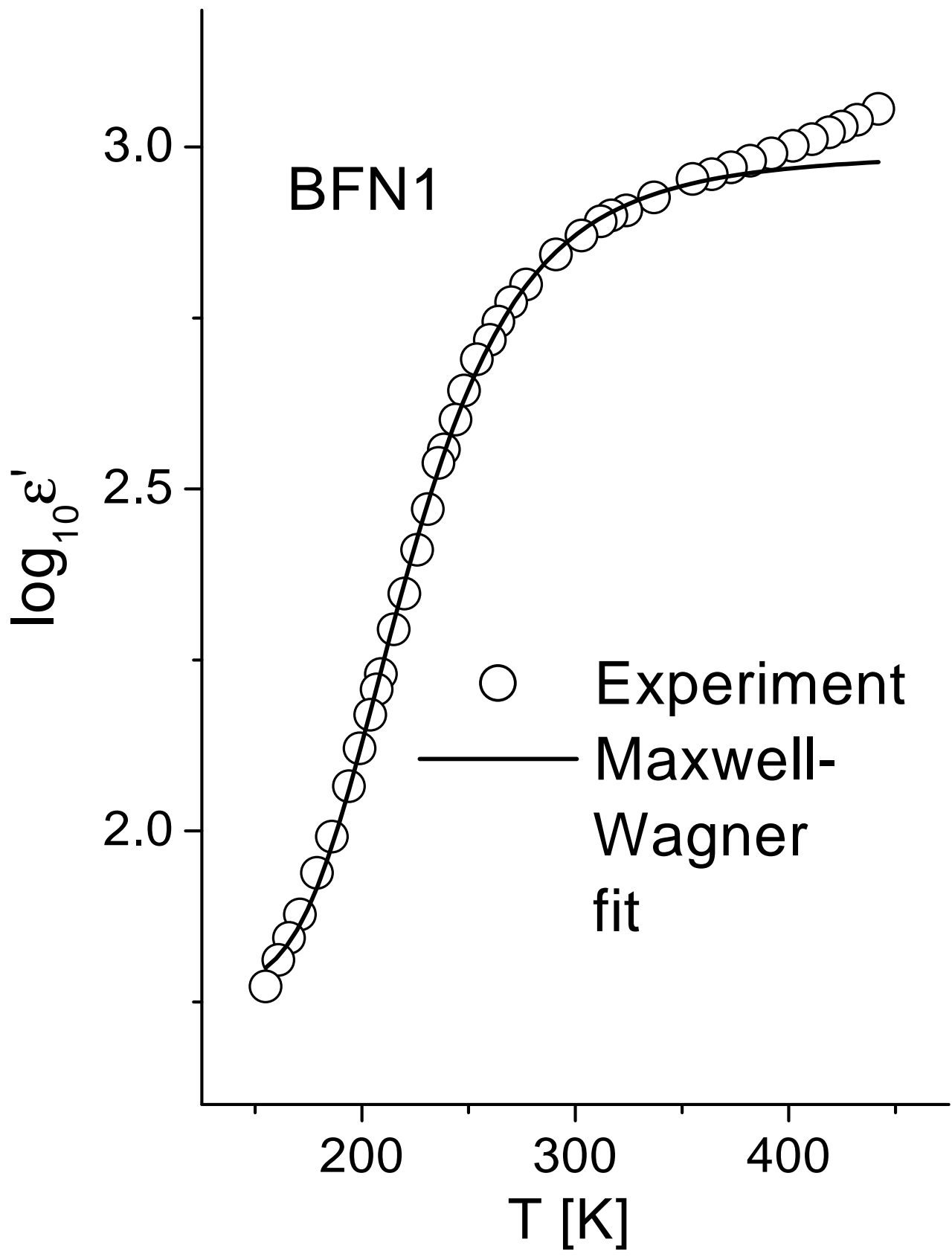
\end{document}